\documentclass{article}
\usepackage{waspaa19,amssymb,amsmath,graphicx,url,times,subfigure,setspace,enumitem,tabularx}
\usepackage{algorithm}
\usepackage[noend]{algpseudocode}
\usepackage{color}
\definecolor{myblue}{rgb}{0.83, 0.89, 0.98}
\definecolor{mypink}{rgb}{0.95, 0.85, 0.8}
\input{def.set}
\usepackage{multirow}
\usepackage{placeins}
\usepackage{caption}
\newcolumntype{L}[1]{>{\raggedright\arraybackslash}p{#1}}
\newcolumntype{C}[1]{>{\centering\arraybackslash}p{#1}}
\newcolumntype{R}[1]{>{\raggedleft\arraybackslash}p{#1}}

\makeatletter
\title{Nearest Neighbor Search-Based Bitwise Source Separation\\Using Discriminant Winner-Take-All Hashing}

\name{Sunwoo Kim, Minje Kim}
\address{Indiana University, Department of Intelligent Systems Engineering, Bloomington, IN, USA 47408\\\texttt{kimsunw@indiana.edu, minje@indiana.edu}}

\begin{document}

\ninept
\maketitle

% To mention
% - maxpool
% - WLOG models optimized
% - do we want IRM, or IBM should be good enough
% - manifold learning

\begin{sloppy}

\begin{abstract}
We propose an iteration-free source separation algorithm based on Winner-Take-All (WTA) hash codes, which is a faster, yet accurate alternative to a complex machine learning model for single-channel source separation in a resource-constrained environment. We first generate random permutations with WTA hashing to encode the shape of the multidimensional audio spectrum to a reduced bitstring representation. A nearest neighbor search on the hash codes of an incoming noisy spectrum as the query string results in the closest matches among the hashed mixture spectra. Using the indices of the matching frames, we obtain the corresponding ideal binary mask vectors for denoising. Since both the training data and the search operation are bitwise, the procedure can be done efficiently in hardware implementations. Experimental results show that the WTA hash codes are discriminant and provide an affordable dictionary search mechanism that leads to a competent performance compared to a comprehensive model and oracle masking.
\end{abstract}


\section{Introduction}
\label{sec:intro}

Numerous data-driven approaches to the source separation problem have attained much improvements in the denoising quality of the enhanced sound. Nonnegative matrix factorization (NMF)-based solutions have shown good performance, which provide not only dimensionality reduction but also an intuitive notion for audio signals \cite{virtanen2007monaural, 4518538}. 
%sun2013universal or virtanen2007monaural?
Though lightweight and effective, NMF algorithms need to pre-specify the number of latent variables that can be found as a hyperparameter. Recently, deep learning approaches have been popular in this domain as well \cite{xu2014experimental}. Fully connected models trained on large sets of mixed spectra have been able to learn complex mappings to their corresponding ideal binary mask (IBM) targets \cite{wang2013towards, grais2014deep}. In addition, recurrent neural networks, which can remember information from previous time frames using hidden states and gating techniques, have further improved denoising performance \cite{mimilakis2018monaural}. Convolutional neural networks \cite{grais2017single} and even Wavenets \cite{rethage2018wavenet} have been successfully explored for source separation as well. 

In the proposed method we formulate the source separation problem as a nearest search process, which is to find the nearest mixture spectrum, and consequently its corresponding IBM vector, in the training set for the given test mixture spectrum. We expedite this tedious search process by a hashing scheme that produces binary codes which enable a bitwise search operation. To this end, we choose the winner-take-all (WTA) hashing algorithm \cite{yagnik2011power}, which has been successfully used in speeding up complex computer vision tasks \cite{dean2013fast}.  Our proposed method provides an affordable, iteration-free, and hardware-friendly bitwise solution to the nearest search problem that finds the source separation solution within the raw training mixture spectra by using the test mixture spectrum as a query. One weakness is the requirement for a large dictionary, but we mitigate it by reducing its size using hashing. 

% In literature, WTA has already been harmonized into topic modeling or NMF-based source separation models. In \cite{kim2015efficient}, the algorithm is an iterative EM procedure and is not a fully bitwise approach which is expensive in that it incurs a linear growth in search space with respect to number of sources. Although in \cite{guo2018bitwise} the E-step was replaced by a search in the hashed spectra, the performance is suboptimal compared to a full EM procedure. It also requires a heavy reliance on the W-disjoint orthogonality \cite{rickard2002approximate}.

% we first construct a pairwise affinity matrix among the training spectra, and then find a discriminant bitstring representation of the training spectra, whose Hamming similarity can approximate the original pairwise similarity. 

\section{Related Works}
\label{sec:format}

\subsection{Manifold preserving source separation}

Instead of learning complex models that can generalize well, the data themselves can act as a more expressive representation. 
In the sparse topic modeling-based source separation \cite{smaragdis2009sparse}, clean source spectra are set as the overcomplete bases for source-specific dictionaries. Incoming mixture spectrum is then decomposed into sparse activations of those predefined bases vectors.
This procedure preserves the manifold of the sources, thereby providing more natural reconstruction of sources.
Nonetheless, manifold preservation necessitates a large source dictionary to extract close estimates. 
A less computationally intensive formulation can be found in the hierarchical latent variable model, where additional latent variables weed out redundant dictionary items during analysis while still retaining the same expressiveness of the data \cite{kim2013manifold}.

A source separation system can employ the $\mathcal{K}$-nearest neighbor ($\mathcal{K}$NN) search outside of the topic modeling or NMF context. For example, a vocal separation method was proposed in \cite{fitzgerald2012vocal}, where the median value of $\mathcal{K}$NN for each mixture frame estimates the background music. However, this approach is an unsupervised algorithm which cannot take advantage of available training data.

In the manifold preserving source separation context, WTA hashing was explored as a fast and low-cost surrogate for searching relevant source candidates in the sparse encoding step \cite{kim2015efficient}. The clean training spectra and source estimates were transformed into binary hash codes, which allowed for an efficient bitwise search to reduce the size of the dictionary. One limitation is the inability of the hash codes to fully reflect the original error function, cross entropy; hence, for a guaranteed performance, a full EM procedure is still required on the reduced dictionary.

To resolve this issue, a fully bitwise voting-based solution was proposed in \cite{guo2018bitwise}. 
Again by applying WTA hashing on the source dictionaries, the algorithm counted the number of matches directly between the hashed mixture and each dictionary to represent the similarity of the mixture to the individual sources. This deemed source estimates unnecessary, and as an additional benefit, it could be performed as a single shot E-step. However, the procedure relies heavily on the W-disjoint orthogonality \cite{rickard2002approximate}, a quality preserved by WTA codes. Furthermore, the separation quality is yet suboptimal due to the randomness in the hash function. We extend this line of work and propose a better-performing, fully bitwise, and iteration-free algorithm with no assumed W-disjoint orthogonality.

\subsection{Winner-Take-All Hashing}

%WTA
% - Hook: Usages in CV and Audio
In image classification \cite{dean2013fast} and audio source separation tasks \cite{kim2015efficient, guo2018bitwise}, WTA hashing has shown its potential in approximated $\mathcal{K}$NN searches. 
% - Why it's good overview
% -- Rank ordering
The core property of WTA is that rank orders of multiple dimensions can preserve the shape of the input vector. 
% it is not the precise values in high-dimensional spaces are not important but their implicit ordering. 
% -- Take max elements which is more important
To this end, WTA approximates the exponentially complex rank order metric by repeatedly sub-sampling $M$ randomly permuted dimensions and recording the position of the winner out of $m$. As each repetition adds more ordering information, the accumulated winner indices form a hash code that holds partial rank orders. 

% - How it works
% -- Formulation
WTA is formulated as follows. Let $\bX \in \mathbb{R}^{N\times D}$ denote $N$ data samples in a $D$-dimensional feature space. Let $\pi_l \in \mathbb{Z}^{M}$ be $M$ permuted indices: $\pi_l=[i^{(l)}_1, i^{(l)}_2, \cdots i^{(l)}_M]$, where $i^{(l)}_m \in \{1,2,\cdots,D\}$. Note that $M<D$. The WTA procedure generates a set of $L$ such permutations $\Pi \in \mathbb{Z}^{L\times M}$. Each permutation $\pi_l$ selects $M$ elements from the input vectors $\bX_{:,\pi_l}\in \mathbb{R}^{n\times m}$, and calculates the positions of the maximum elements, which forms $l$-th integer hash code $\bcalX_{:,l}=\argmax_m  \bX_{:,\pi_i(m)}$. By repeating this process for all $L$ permutations, the retrieved $L$ integers per sample form $N$ hash codes $\bcalX^\Pi \in \mathbb{Z}^{N\times L}$. For example, for input vectors of $D=4$, suppose $L=2$ and $M=3$ such that $\pi_1 = [4,1,2]$ and $\pi_2 = [3,4,2]$. Then, the WTA hash codes for an input $\bX_{n,:} = [6.6, 2.2, 4.4, 3.3]$ is $\bcalX_{n,:} = [2,1]$ as $\bX_{n,\pi_1}=[3.3, \mathbf{6.6}, 2.2]$ and $\bX_{n,\pi_2}=[\mathbf{4.4},3.3,2.2]$.

There are several benefits of employing WTA. First, it nonlinearly transforms the data samples into binary features preserving the rank correlation of the original representation.
% Hamming distance approximate original
% --- LSH property and since sparse binary vectors it is fast similarity search
Also, as with locality sensitive hashing \cite{gionis1999similarity}, the encodings for similar data points have higher probability of collision.
% Fast Comparison instead of metric distance
Furthermore, the Hamming metric can expedite comparisons and the similarity search.
% --- Robust to permutations (rank corrrelation measures/ ordinal representations/ partial order statistics)
Lastly, the partial rank order statistics encoded in the bistrings share the benefits of rank correlation measures such as robustness to additive noise. 

\subsection{Kernel-based source separation}

\begin{figure}[!tbp]
    \centering
\subfigure[Original]{\includegraphics[width=0.49\columnwidth]{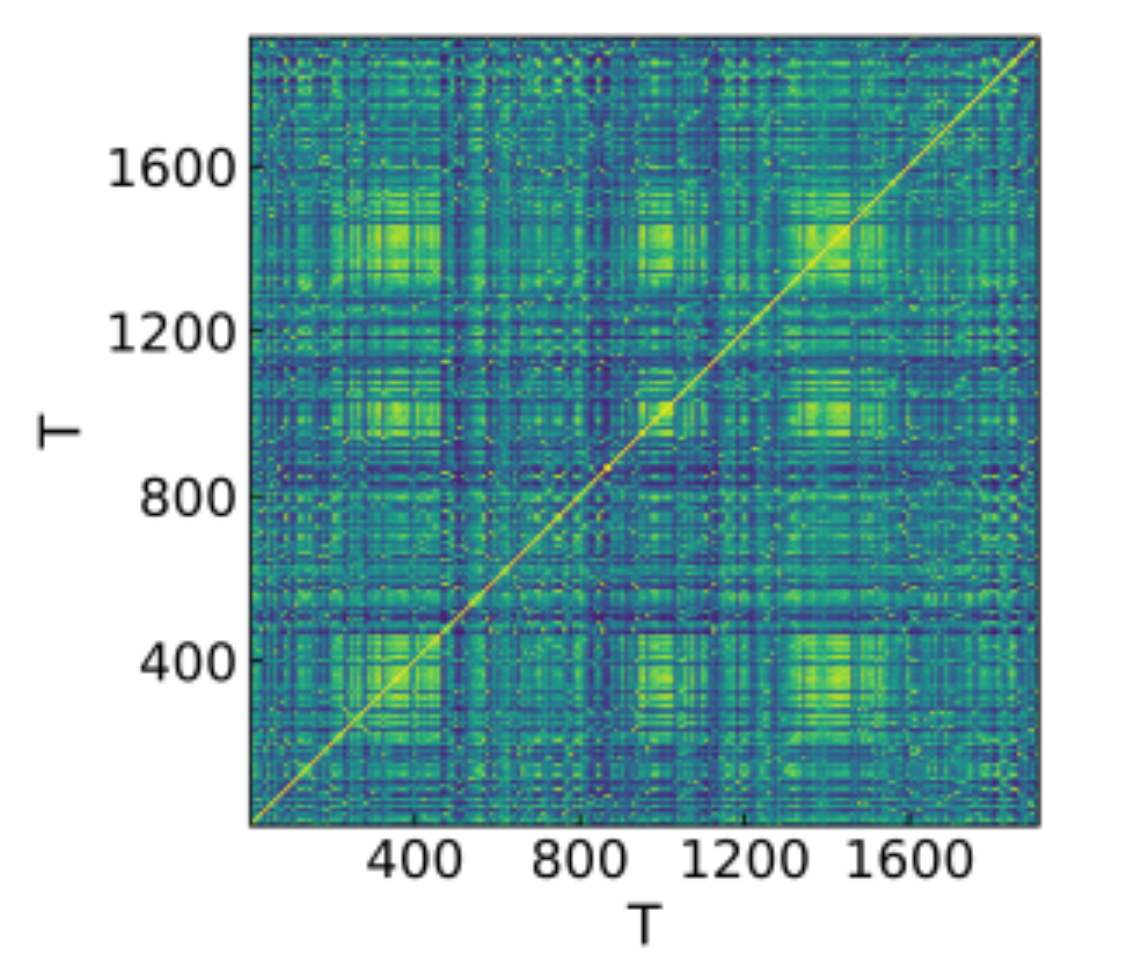}}    
\subfigure[WTA hash codes]{\includegraphics[width=0.49\columnwidth]{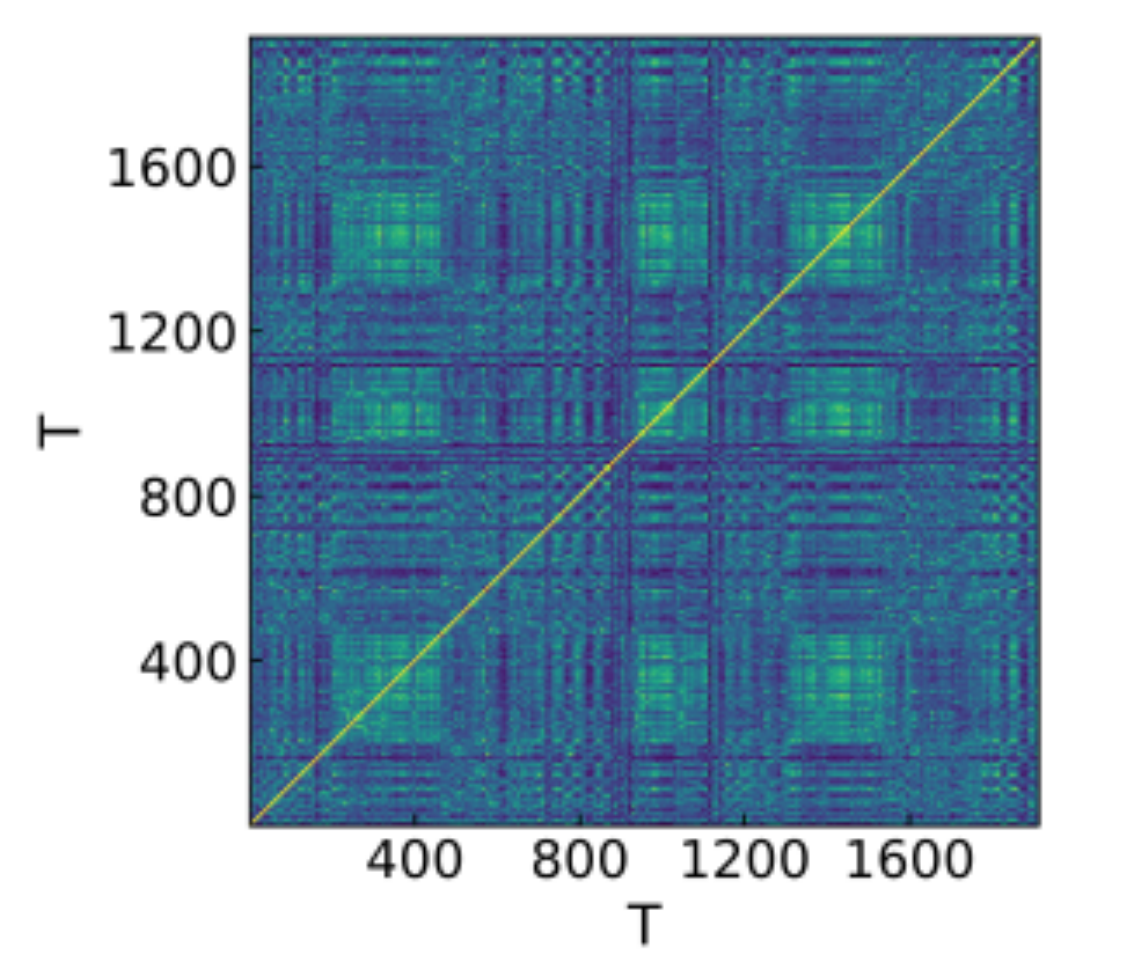}}
  \caption{Self-affinity matrices of original time-frequency bins and WTA hash codes.}
  \label{fig:affinity}
\end{figure}

Although we rely on the randomly generated permutation table $\Pi$, we wish that it leads to discriminant binary embeddings that preserve pairwise similarity among original data samples. Finding embeddings that preserve the semantic similarity is a popular goal in many disciplines. 
In natural language processing, Word2Vec \cite{mikolov2013distributed} or GloVe \cite{pennington2014glove} methods use pairwise metric learning to retrieve a distributed contextual representation that retains complex syntactic and semantic relationships within documents. 
Another model that trains on similarity information is the Siamese networks, which learn to discriminate a pair of examples \cite{bromley1994signature}. 

Utilizing similarity information has been explored in the source separation community by posing denoising as a segmentation problem in the time-frequency plane with an assumption that the affinities between time-frequency regions of the spectrogram could condense complex auditory features together \cite{bach2006learning}. Inspired by studies of perceptual grouping \cite{cooke2001auditory}, in \cite{bach2006learning} local affinity matrices were constructed out of cues specific to that of speech. Then, spectral clustering segments the weighted combination of similarity matrices to unmix speech mixtures. 
On the other hand, deep clustering learned a neural network encoder that produces discriminant spectrogram embeddings, whose objective is to approximate the ideal pairwise affinity matrix induced from IBM \cite{hershey2016deep}. 

Our proposed method also learns a transformation function as in deep clustering, but in the form of a WTA hash function, which still approximates the ground-truth affinity matrix in the binary embedding space. Figure \ref{fig:affinity} illustrates this self-affinity preserving quality of the WTA hash codes. Also, our method predicts an IBM vector per frame rather than attempting to segment spectrogram bins.

\section{WTA Hashing for $\mathcal{K}$NN Source Separation}
\label{sec:model}

\subsection{$\mathcal{K}$NN search-based source separation}

We keep consistent with preserving the manifold by maintaining the excessively many training examples as the search space and finding only $\calK$NN to infer the mask. We assume that if the mixture frames are similar, the sources in the mixture as well as their IBMs must also be similar. We also assume that the average of IBMs of $\calK$NN is a good estimate of the ideal ratio mask (IRM).

Let $\bH \in \mathbb{R}^{T\times D}$ be the feature vectors from $T$ frames of training mixture examples. As our training examples are the mixture signals of the sources of interest, $T$ can be a potentially very large number as it grows with the number of sources. Out of many potential choices, we are interested in short-time Fourier transform (STFT) and mel-spectra for feature extraction. For example, if $\bH$ is from STFT on the training mixture signals, $D$ equals the number of subbands $F$ in each spectrum, while for mel-spectra $D<F$. We also prepare their corresponding IBM matrix, $\bY \in \mathbb{Z}^{T\times F}$, whose dimension $F$ matches that of STFT. For a feature vector of an incoming test mixture frame $\bx\in\mathbb{R}^{D}$, our goal is to estimate a denoising mask, $\hat{\by} \in \mathbb{R}^{F}$, to recover the source by masking, $\hat{\by}\odot\bx$, for which $\bx$ should be with full $F$ complex Fourier coefficients.

\begin{algorithm}[t]
\caption{$\mathcal{K}$NN source separation}\label{alg:knn}
\begin{algorithmic}[1]
\State Input: $\bx$, $\bH$ \Comment{A test mixture vector and the dictionary}
\State Output: $\hat{\by}$ \Comment{A denoising mask vector}
\State Initialize an empty set $\calN=\varnothing$ and $\mathcal{A}_\text{min}=0$
\For{$t \leftarrow 1$ to $T$}
    \If{$\calS_{\text{cos}}(\bx, \bH_{t,:})>\mathcal{A}_\text{min}$}
        \State{Replace the farthest neighbor index in $\calN$ with $t$}
        \State{Update $\mathcal{A}_\text{min}\leftarrow \min_{k\in \calN} \calS_{\text{cos}}(\bx, \bH_{k,:})$}
        % \Comment{Update the farthest distance to the current neighbors}
    \EndIf 
\EndFor
\State{return $\hat{\by} \leftarrow \frac{1}{\calK} \sum_{k \in \mathcal{N}}\bY_{k,:}$}
\end{algorithmic}
\end{algorithm}

Algorithm \ref{alg:knn} describes the $\calK$NN source separation procedure. We use notation $\mathcal{S}_{\text{cos}}$ as the affinity function, e.g., the cosine similarity function. For each frame $\bx$ in the mixture signal, we find the $\mathcal{K}$ closest frames in the dictionary (line 4 to 7), which forms the neighborhood set $\calN=\{\tau_1, \tau_2, \cdots, \tau_\calK\}$. Using them, we find the corresponding IBM vectors from $\bY$ and take their average (line 8). 

\textbf{Complexity:} The search procedure is non-iterative but requires a linear scan of all frames in $\bH$, giving $O(T)$. This procedure is restrictive since $T$ needs to be large for good source separation. In the next section, we apply WTA hashing to convert $\bH$ into integer values $\bcalH\in \mathbb{Z}^{L\times T}$ to perform the search in a bitwise fashion.

\begin{figure}
    \centering
    \includegraphics[width=\columnwidth]{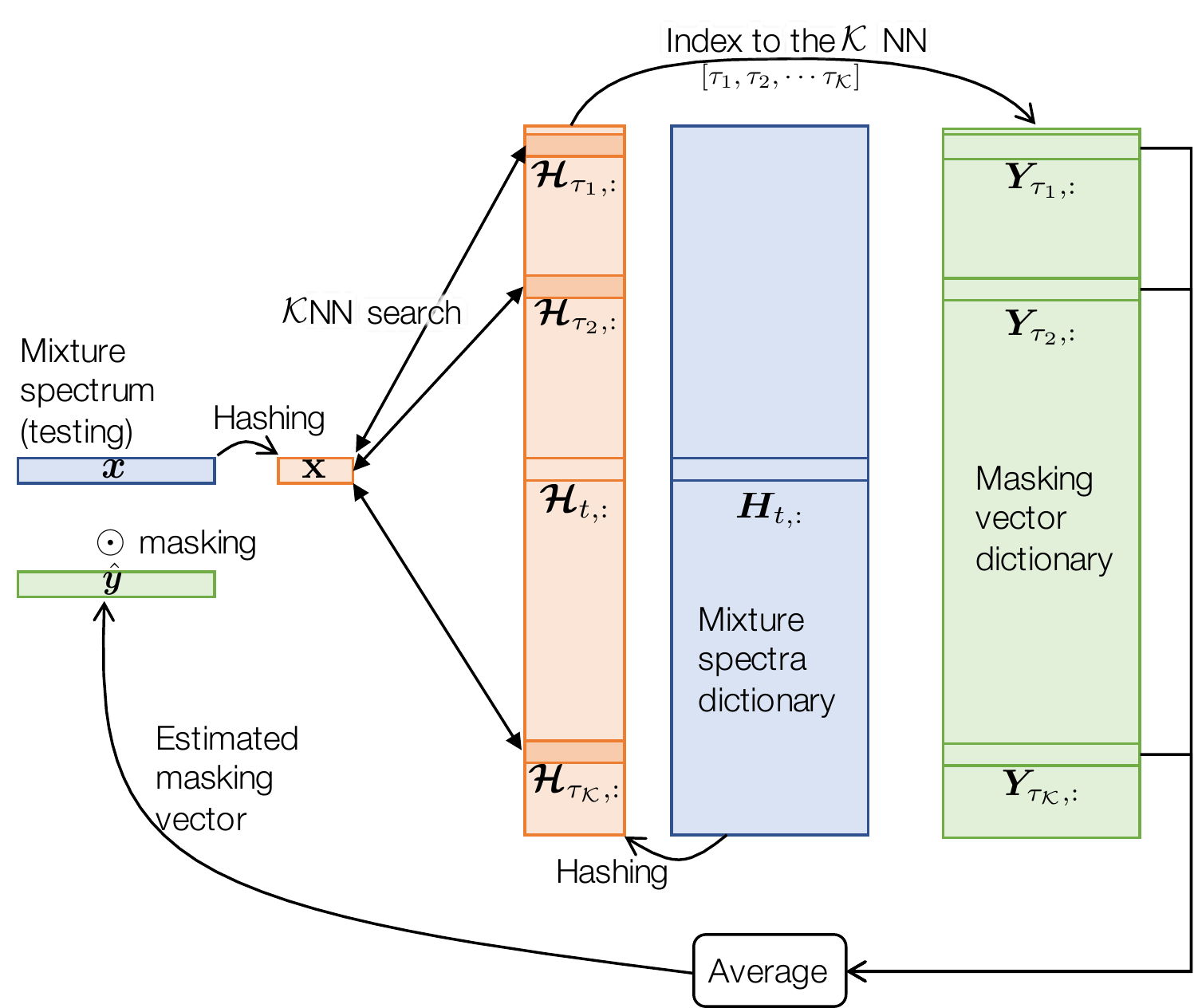}
    \caption{The $\mathcal{K}$NN-based source separation process using WTA hash codes.}
    \label{fig:system}
\end{figure}

\subsection{$\mathcal{K}$NN-based source separation on WTA hash codes}\label{sec:wtaknn}
We can expedite Algorithm \ref{alg:knn} using hashed spectra and the Hamming similarity between them. To this end, we first generate the $L$ random permutations $\Pi \in \mathbb{R}^{L\times M}$, which is used to convert $\bH$ and $\bx$ to obtain $\bcalH^{\Pi} \in \mathbb{Z}^{L\times T}$ and $\mathbf{x}^{\Pi} \in \mathbb{Z}^{L\times 1}$, respectively. By having them as the new feature representations, we apply Algorithm \ref{alg:knn}, but this time with Hamming similarity as the similarity function that counts the number of matching integers: $\calS_\text{Ham}(\ba, \bb)=\sum_l \calI (\ba_l, \bb_l)/L$, where $\calI(x,y)=1 \text{ iff } x=y$. The other parts of the algorithm are the same. Figure \ref{fig:system} describes the source separation process using $\calK$NN searches on hash codes.

\textbf{Complexity:} Since the same Algorithm \ref{alg:knn} is used, the time complexity is still $O(T)$. Nonetheless, the procedure is significantly accelerated since the binarized feature vectors allow the Hamming similarity calculation to be done through bitwise AND operations. In addition, the spatial complexity reduces significantly from $O(64D)$ to $O(\lceil\log_2 M \rceil L)$, where 64 and $\lceil\log_2 M \rceil$ are the number of bits to store an element in $\bH$ (double precision) and $\bcalH$, respectively. Our experiments choose $M\leq 2$, and $L<D$.

Some degradation in performance is expected due to quantization error. Theoretically, the asymptotic behavior as $L \rightarrow \infty$, the hash codes closely approximates the full rank order metric, but there is a mismatch between the full rank order metric and the choice of similarity function in the original feature space, e.g., $\calS_\text{cos}$.  Hence, increasing $L$ does not always guarantee the best result.
Another problem with this approach is the randomness in computing $\Pi$, whose quality as a hash function fluctuates. A more consistent result can be achieved by repeating and averaging results from different $\Pi$ tables, while the repetition will multiply the size of $L$.

\begin{table*}[hbtp!]
\parbox{.67\textwidth}{
\caption{Speech denoising performance of the proposed WTA source separation model compared with oracle and plain $\mathcal{K}$NN procedure.} \label{tab:table}
\resizebox{.68\textwidth}{!}{
\begin{tabular}{|R{0.8cm}|R{0.8cm} R{0.8cm} R{0.8cm}|R{0.8cm} R{0.8cm} R{0.8cm}|R{0.8cm} R{0.8cm} R{0.8cm} |R{0.8cm} R{0.8cm} R{0.8cm}|}
\hline
 Noise & \multicolumn{3}{c|}{SDR} & \multicolumn{3}{c|}{SIR} & \multicolumn{3}{c|}{SAR} & \multicolumn{3}{c|}{STOI} \\
 \cline{2-13} 
types& 
Oracle & $\mathcal{K}$NN & WTA & 
Oracle & $\mathcal{K}$NN & WTA & 
Oracle & $\mathcal{K}$NN & WTA &
Oracle & $\mathcal{K}$NN & WTA \\
\hline
1 & 17.90 & 11.96 & 11.77 & 27.01 & 21.18 & 20.42 & 18.51 & 12.55 & 12.45 & 0.96 & 0.82 & 0.81 \\
2 & 12.77 & 6.58 & 5.34 & 22.48 & 18.48 & 17.00 & 13.28 & 6.93 & 5.74 & 0.94 & 0.74 & 0.71 \\
3 & 21.02 & 17.27 & 17.01 & 30.87 & 29.63 & 29.80 & 21.54 & 17.59 & 17.30 & 0.99 & 0.95 & 0.95 \\
4 & 16.41 & 10.75 & 10.06 & 28.86 & 21.65 & 21.85 & 16.67 & 11.16 & 10.42 & 0.94 & 0.83 & 0.82 \\
5 & 19.23 & 13.52 & 12.77 & 26.71 & 22.61 & 22.11 & 20.26 & 14.19 & 13.36 & 0.97 & 0.88 & 0.87 \\
6 & 17.21 & 11.32 & 11.28 & 27.10 & 19.50 & 19.25 & 17.69 & 12.12 & 12.01 & 0.97 & 0.85 & 0.84 \\
7 & 14.01 & 7.56 & 7.10 & 23.85 & 17.41 & 16.33 & 14.53 & 8.13 & 7.64 & 0.96 & 0.78 & 0.77 \\
8 & 16.69 & 12.46 & 11.37 & 28.65 & 28.57 & 29.15 & 16.99 & 12.58 & 11.25 & 0.94 & 0.87 & 0.86 \\
9 & 15.26 & 10.43 & 9.74 & 24.92 & 23.38 & 23.42 & 15.77 & 10.68 & 9.74 & 0.93 & 0.79 & 0.77 \\
10 & 11.94 & 6.77 & 5.56 & 20.98 & 18.32 & 17.49 & 12.56 & 7.16 & 5.74 & 0.91 & 0.68 & 0.65 \\
\hline
%Avg & - & - & - & - & - & - & - & - & - & - & - & - \\
%\hline
\end{tabular}
}
}
\hfill
\parbox{.3\textwidth}{
\caption{A comparison of different models.} \label{tab:comp}
\resizebox{.29\textwidth}{!}{
\begin{tabular}{|c || r | r | r|} 
      \hline
  Systems & SDR & SIR & SAR \\ 
  \hline
  Oracle  &16.24 & 26.14& 16.78\\
  \hline
  $\calK$NN on mel & 10.86 & 20.07 & 11.31\\
 \hline
 $\calK$NN-WTA & 10.20 & 21.68 & 10.56 \\ 
 \hline
 KL-NMF (Male) & 10.23 & - & - \\
 \hline
 USM (Male) & 10.41 & - & - \\ 
 \hline
 WTA on E-step & 7.47 & 10.03 & 9.35 \\
 \hline
\end{tabular}
}\vspace{0.8in}}

\end{table*}

\section{Experiments}
\label{sec:experiment}
\subsection{Experimental setups}
For the experiment, we randomly subsample 16 speakers for training and 10 speakers for testing from the TIMIT corpus with a gender balance, where each speaker is with ten short recordings of various utterances with a 16kHz sampling rate. Each utterance is mixed with ten different non-stationary noise sources with 0 dB signal-to-noise ratio (SNR), namely $\{$birds, casino, cicadas, computer keyboard, eating chips, frogs, jungle, machine guns, motorcycles, ocean$\}$ \cite{duan2012online}. 
For each noise type, we have 1,600 training utterances consisting of approximately 15,000 frames to build our mixture dictionary and ten query utterances.
%For each noise type, we use all mixture utterances from training speakers as training examples and sample at random one utterance each from the unseen test speakers as testing examples. In total, we have 1,600 training utterances consisting of approximately X frames to build our mixture dictionary and ten \minje{100?} query utterances. 
We apply a short-time Fourier transform (STFT) with a Hann window of 1024 and hop size of 512 and transform these into mel-spectrograms. For evaluation of the final results, we used signal-to-distortion ratio (SDR), signal-to-interference ratio (SIR), signal-to-artifact ratio (SAR) \cite{vincent2006performance}, and short-time objective intelligibility (STOI) \cite{taal2010short}.

We compare three systems we implement, and three other kinds from the literature:

\vspace{-0.04in}
\begin{itemize}[leftmargin=0in]\setlength{\itemindent}{.15in}
    \item \textbf{\em Oracle IRM}: We apply the ground-truth IRM to the test signal to calculate the performance bound of the source separation task.\vspace{-0.04in}
    \item \textbf{\em $\calK$NN on the original spectra}: For each given test spectrum we can find the best matches from the dictionary by using $\calS_\text{cos}$ as the similarity metric (Algorithm \ref{alg:knn}). Hashing-based technics try to catch up this performance. \vspace{-0.04in}
    \item \textbf{\em $\calK$NN on the WTA hash codes}: Performs $\calK$NN separation using $\calS_\text{Ham}$ (Section \ref{sec:wtaknn}). \vspace{-0.04in}
    \item \textbf{\em KL-NMF}: An NMF-based model that learns noise- and speaker-specific dictionaries. It is fully supervised, while our $\calK$NN models are specific only to noise types. We are based on the experimental results reported in \cite{sun2013universal}.
    \item \textbf{\em Universal speech model (USM)}: USM is another NMF-based fully supervised model that uses 20 speaker-specific dictionaries, but only a few of them are activated during the test time \cite{sun2013universal}. 
    % \item \textbf{\em Online PLSA}:
    \item \textbf{\em Bitwise E-step using WTA}: Another variant that uses WTA process to replace the posterior estimation (E-step) in topic modeling \cite{guo2018bitwise}. This one is based on a large speech dictionary from 32 randomly chosen training speakers, while the noise type is known. 
    % \item \textbf{\em $\calK$NN on the proposed dWTA hash codes}: Uses the proposed dWTA hash codes. \vspace{-0.04in}
\end{itemize}

% The experiment is conducted by first incrementally constructing the permutation table $\Pi^*$ as in Algorithm \ref{alg:dwta}, which results in the hashed spectra that approximates the original pair-wise similarity. The permutations are applied on both the dictionary and noisy utterance from unseen speakers of the same noise type. With the set of permutations, the noisy signal is processed using the $\mathcal{K}$ source separation procedure.

% The proposed dWTA method has three major parameters $\mathcal{K}$, $L$, and $M$. To contain the search space, we fix the value $L=100$. This is based on the asymptotic property of $L$ that as $L\rightarrow \infty$ the $\bS^{\calH}$ will converge to $\bS^{\bH}$, from which we can already infer an increase in performance from additional permutations. Also, finding permutations that further improve the divergence can take much longer. Hence, we search for the best $\mathcal{K}$ and $M$ parameters with a similar approach as \cite{kim2015efficient} and \cite{guo2018bitwise}. This is described in more detail in the later section. 

\subsection{Experimental results and discussion}

\begin{itemize}[leftmargin=0in]\setlength{\itemindent}{.15in}
\item \textbf{\em WTA parameters}: We explore $\mathcal{K}$ and $M$ for our WTA source separation algorithm while fixing $L=100$. $M$ is the number of samples in comparison to find the winner in each permutation. Larger $M$ value can exploit the distribution of the input vector more to some degree, but a too large $M$ is detrimental as it breaks down the locality assumption. $\mathcal{K}$ is the number of nearest neighbor frames we search from the dictionary. Larger $\mathcal{K}$ would provide more examples for the IRM estimation. However, more neighbors do not always correlate with better source separation performance, similarly to the $\mathcal{K}$NN classification case. Furthermore, $\mathcal{K}$ and $M$ define the computational complexity of the system as discussed in Section \ref{sec:model}. Figure \ref{fig:results} (a) illustrates a grid search result on $\mathcal{K}$ and $M$ from the WTA-based $\calK$NN source separation. For the given combination, we perform separation on all ten noise types and take the average. $\mathcal{K}=5$ and $M=6$ are the best combination, giving peak average performance of 10.20 dB. 

%with performance being X dB higher than the fully bitwise method that even searched for parameters for individual noise types \cite{guo2018bitwise}, and 2dB higher than a partially bitwise method \cite{kim2015efficient}.

We present a closer look at the best parameters in Table \ref{tab:table}. The results show the source separation performance over all ten noises of the oracle, $\mathcal{K}$NN, and WTA. 
The performance of WTA catches up to that of $\mathcal{K}$NN for all metrics and obtains higher SIR values for certain noise types. For noise type 8, the WTA method even outperforms the oracle in terms of SIR. 
This shows that the $\mathcal{K}$NN procedure works well in the hashed space, which is expected from the property of WTA that claims that the ranking correlation measures are preserved with the Hamming metric. 
Some decrease in performance, however, is shown and this is expected since the hashing procedure incurs a quantization error. 
\vspace{-0.04in}
    
\end{itemize}

\textbf{Spectrogram format:} Some time-frequency bins in a STFT spectrogram, especially in the high frequency, are with minuscule values. The mel-spectrogram, on the other hand, is on a logarithmic scale with lower frequency resolution in the high frequency. Therefore, we expect that WTA hashing on mel-spectra is based on comparisons involving more lower frequency bins, thus making winner indices more representative. Furthermore, mel-spectra is with much smaller dimensionality, a property that makes WTA hashing easier to preserve locality. The described effect is illustrated in Figure \ref{fig:results} (b). Note that for certain noises, WTA on a STFT spectrogram performs much worse, and even returns a negative SDR value. 

\begin{figure}[!tbp]
    \centering
\subfigure[]{\includegraphics[width=0.38\columnwidth]{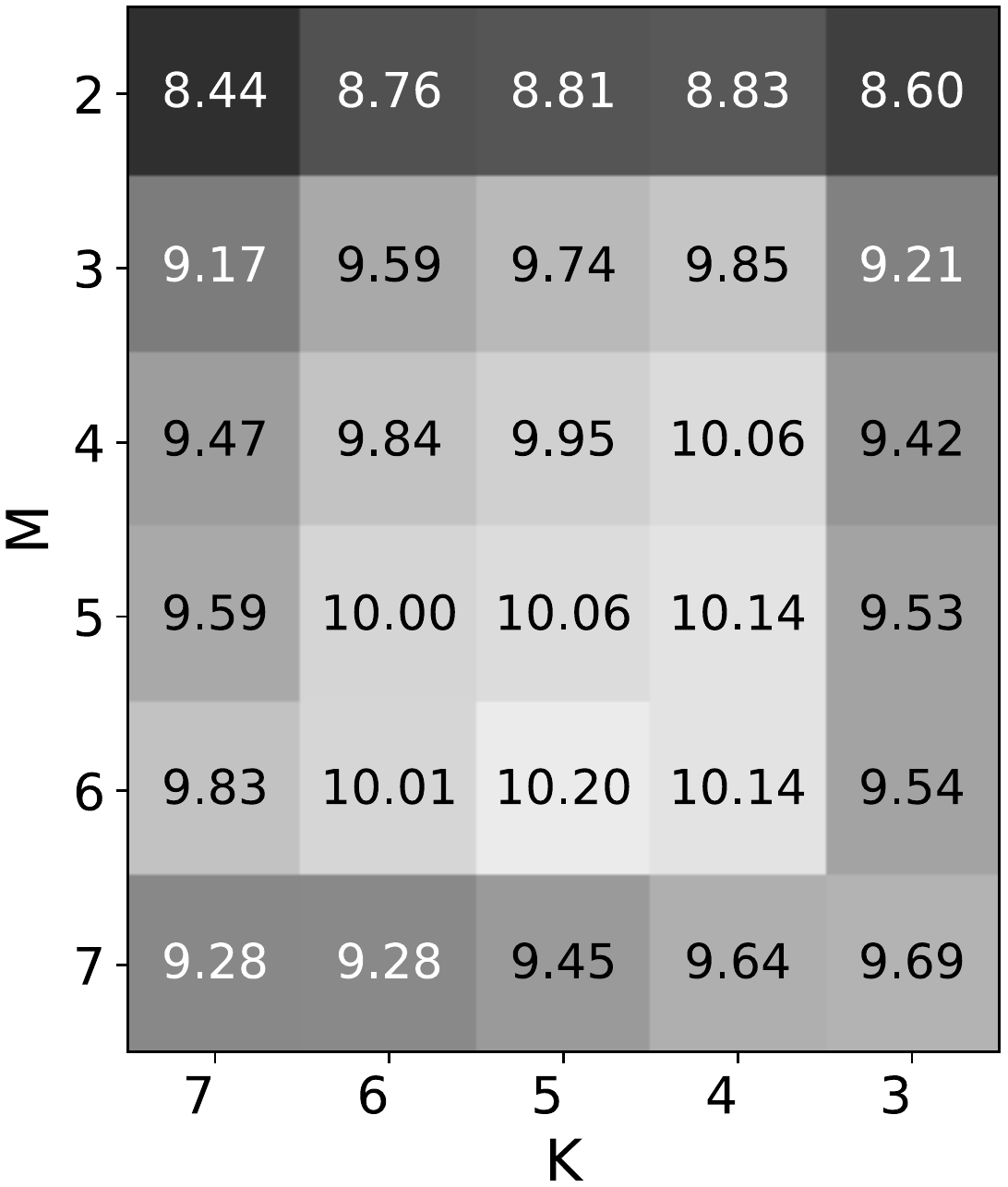}}    
\subfigure[]{\includegraphics[width=0.61\columnwidth]{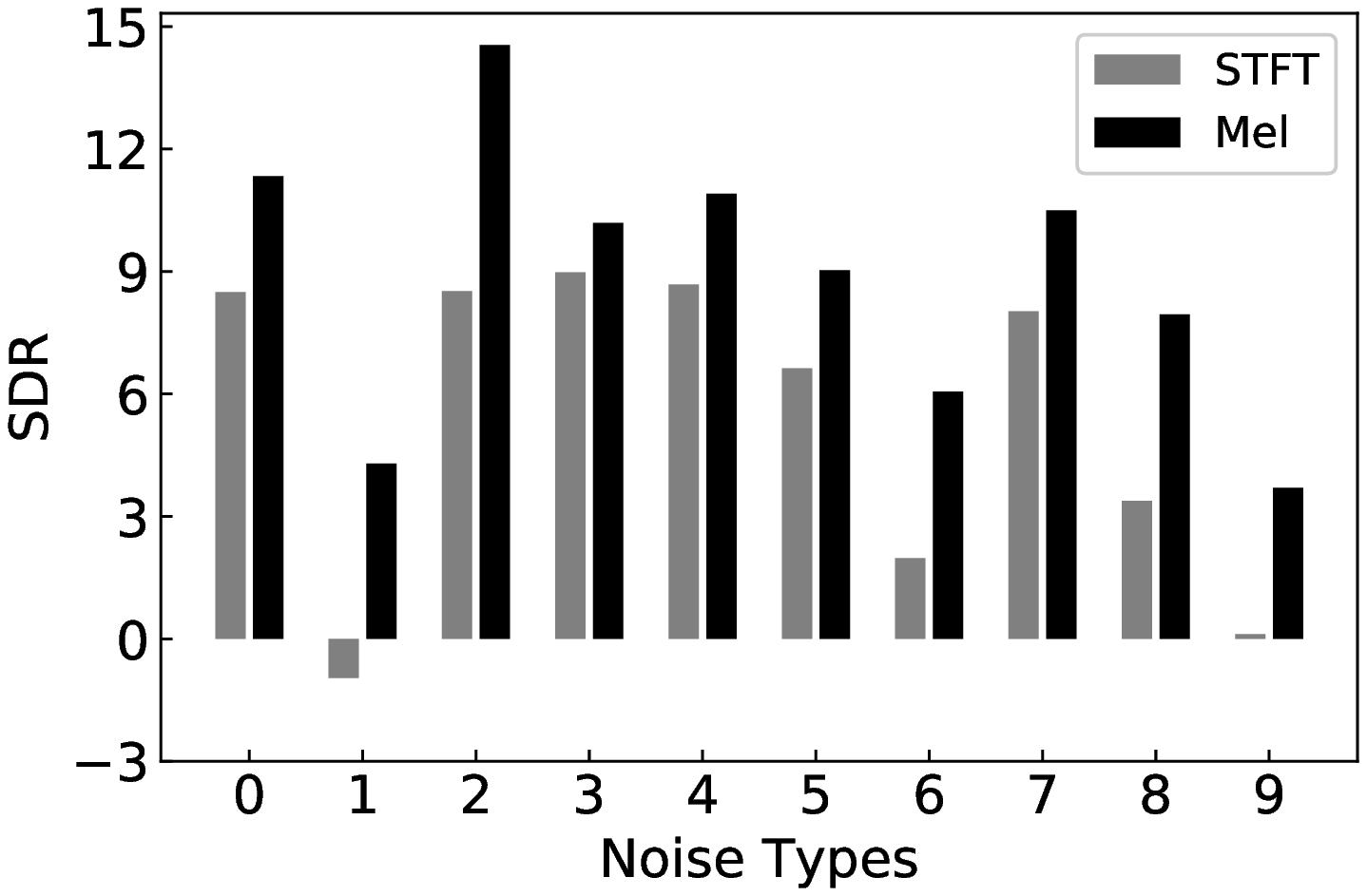}}
  \caption{(a) Comparison of source separation on STFT and Mel spectra in terms of SDR (b) Results of WTA on various $\mathcal{K}$ and $M$.}
  \label{fig:results}
\end{figure}

\textbf{Comparison with other dictionary based methods:} 

Table \ref{tab:comp} shows the results of the proposed method along with those of other dictionary based methods that use KL-NMF, USM, and the bitwise posterior estimation using WTA (WTA on E-step). 
Although each of the methods used different speaker sets, they were all tested using the same noise types in similar scenarios. KL-NMF and USM learns 10 basis vectors from the speakers, although KL-NMF assumes the known speaker identity and USM does not. Both of them are only on male speakers. WTA on E-step uses gender-balanced 64 speakers' spectra as its large dictionary. Our proposed WTA method achieves either much higher or at least similar performance against all the mentioned methods. Additionally, $\mathcal{K}$NN-WTA is iteration-free. Although it requires a linear scan for the nearest neighbor search, it can be done efficiently with bitwise operations and avoid accruing additional run time.

Comparing against KL-NMF, a real-valued model that assumes the speaker identity to be known, our method on unseen test speaker sets does slightly worse (10.20 versus 10.23). Another NMF based approach, USM, utilizes a much larger dictionary and block sparsity as a regularizer, and performs slightly better than $\mathcal{K}$NN-WTA (10.20 versus 10.41). Finally, we compare the fully bitwise models together, $\mathcal{K}$NN-WTA and WTA on E-step. As the proposed $\mathcal{K}$NN-WTA is free from EM-based estimation, it benefits directly from the abundance of the data and outperforms the topic modeling-based bitwise model with a large margin (10.20 versus 7.47).

It should be noted that $\mathcal{K}$NN-WTA is performed in the fully bitwise domain, but it still competes with the other real-valued NMF-based models. Comparing against just $\mathcal{K}$NN on mel-spectra, it can be seen that our method incurred some penalty from the hashing process; however, the loss is minimal and acceptable given the massive reduction in computational cost from using bitwise operations. Furthermore, the hashing based approach is far more practical to deploy in extreme environments where resource is strictly constrained.

% L value
% \textbf{Incremental addition of permutations:} We next delve into the effects of incremental addition of $L$ individual permutation vectors. Fig. \ref{fig:line} shows the average cross entropy error of each $R=5$ random samples as $\pi$ are added. For visual purposes, the illustration is based on an example run with a single noise type. It can be seen for all $R$ random samples with $L/R$ permutations the convergence is similar, and the error is much lower than that of a plain WTA with $L$ permutations on the entire sample. This illustrates the effective preservation of the self-affinity matrix in dWTA. 

% Metrics
% -- SDR
% -- TODO: STOI also
% Compare with other models FC, CNN, RNN, deep NMF, 

% Future work: 
% Orthogonal MP
% NN
% Adaboost
% low rank approximation to the affinity matrix

% Say how WTA can be good but there's randomness so gotta try many tiesmes or doin't know but our procedure is guaranteed

% Effects of variance on distribution of pairwise distances and ssceptibiltiy to noise this way

%RS may amplify performance?
% Ours is speaker non specific

\section{Conclusion}
\label{sec:conclude}
In this paper, we proposed a fully bitwise nearest neighbor based algorithm for source separation. The WTA source separation model generates permutations and transforms the dictionary and noisy signal into a hashed feature space in which the original rank correlation is preserved with the Hamming metric. This not only allows compression into binary bit strings, but it also allows the use of bitwise operations for the Hamming distance, further reducing computation. Hence, search for $\mathcal{K}$ nearest neighbors can be done efficiently from which to estimate the IRM. For good parameters, $\mathcal{K}$NN-WTA performs well for the speech denoising job reaching SDR values above 10dB. Future directions to explore are minimizing the quantization error introduced from hashing by exploring self-affinity matrices of the hashed and original for better hashed representation.

\bibliographystyle{IEEEtran}
\bibliography{refs19}

\begin{thebibliography}{10}
\providecommand{\url}[1]{#1}
\def\UrlFont{\rmfamily}
\providecommand{\newblock}{\relax}
\providecommand{\bibinfo}[2]{#2}
\providecommand\BIBentrySTDinterwordspacing{\spaceskip=0pt\relax}
\providecommand\BIBentryALTinterwordstretchfactor{4}
\providecommand\BIBentryALTinterwordspacing{\spaceskip=\fontdimen2\font plus
\BIBentryALTinterwordstretchfactor\fontdimen3\font minus
  \fontdimen4\font\relax}
\providecommand\BIBforeignlanguage[2]{{%
\expandafter\ifx\csname l@#1\endcsname\relax
\typeout{** WARNING: IEEEtran.bst: No hyphenation pattern has been}%
\typeout{** loaded for the language `#1'. Using the pattern for}%
\typeout{** the default language instead.}%
\else
\language=\csname l@#1\endcsname
\fi
#2}}

\bibitem{virtanen2007monaural}
T.~Virtanen, ``Monaural sound source separation by nonnegative matrix
  factorization with temporal continuity and sparseness criteria,'' \emph{IEEE
  transactions on audio, speech, and language processing}, vol.~15, no.~3, pp.
  1066--1074, 2007.

\bibitem{4518538}
K.~W. {Wilson}, B.~{Raj}, P.~{Smaragdis}, and A.~{Divakaran}, ``Speech
  denoising using nonnegative matrix factorization with priors,'' in \emph{2008
  IEEE International Conference on Acoustics, Speech and Signal Processing},
  March 2008, pp. 4029--4032.

\bibitem{xu2014experimental}
Y.~Xu, J.~Du, L.-R. Dai, and C.-H. Lee, ``An experimental study on speech
  enhancement based on deep neural networks,'' \emph{IEEE Signal processing
  letters}, vol.~21, no.~1, pp. 65--68, 2014.

\bibitem{wang2013towards}
Y.~Wang and D.~Wang, ``Towards scaling up classification-based speech
  separation,'' \emph{IEEE Transactions on Audio, Speech, and Language
  Processing}, vol.~21, no.~7, pp. 1381--1390, 2013.

\bibitem{grais2014deep}
E.~M. Grais, M.~U. Sen, and H.~Erdogan, ``Deep neural networks for single
  channel source separation,'' in \emph{Acoustics, Speech and Signal Processing
  (ICASSP), 2014 IEEE International Conference on}.\hskip 1em plus 0.5em minus
  0.4em\relax IEEE, 2014, pp. 3734--3738.

\bibitem{mimilakis2018monaural}
S.~I. Mimilakis, K.~Drossos, J.~F. Santos, G.~Schuller, T.~Virtanen, and
  Y.~Bengio, ``Monaural singing voice separation with skip-filtering
  connections and recurrent inference of time-frequency mask,'' in \emph{2018
  IEEE International Conference on Acoustics, Speech and Signal Processing
  (ICASSP)}.\hskip 1em plus 0.5em minus 0.4em\relax IEEE, 2018, pp. 721--725.

\bibitem{grais2017single}
E.~M. Grais and M.~D. Plumbley, ``Single channel audio source separation using
  convolutional denoising autoencoders,'' in \emph{2017 IEEE Global Conference
  on Signal and Information Processing (GlobalSIP)}.\hskip 1em plus 0.5em minus
  0.4em\relax IEEE, 2017, pp. 1265--1269.

\bibitem{rethage2018wavenet}
D.~Rethage, J.~Pons, and X.~Serra, ``A wavenet for speech denoising,'' in
  \emph{2018 IEEE International Conference on Acoustics, Speech and Signal
  Processing (ICASSP)}.\hskip 1em plus 0.5em minus 0.4em\relax IEEE, 2018, pp.
  5069--5073.

\bibitem{yagnik2011power}
J.~Yagnik, D.~Strelow, D.~A. Ross, and R.-s. Lin, ``The power of comparative
  reasoning,'' in \emph{2011 International Conference on Computer
  Vision}.\hskip 1em plus 0.5em minus 0.4em\relax IEEE, 2011, pp. 2431--2438.

\bibitem{dean2013fast}
T.~Dean, M.~A. Ruzon, M.~Segal, J.~Shlens, S.~Vijayanarasimhan, and J.~Yagnik,
  ``Fast, accurate detection of 100,000 object classes on a single machine,''
  in \emph{Proceedings of the IEEE Conference on Computer Vision and Pattern
  Recognition}, 2013, pp. 1814--1821.

\bibitem{smaragdis2009sparse}
P.~Smaragdis, M.~Shashanka, and B.~Raj, ``A sparse non-parametric approach for
  single channel separation of known sounds,'' in \emph{Advances in neural
  information processing systems}, 2009, pp. 1705--1713.

\bibitem{kim2013manifold}
M.~Kim and P.~Smaragdis, ``Manifold preserving hierarchical topic models for
  quantization and approximation,'' in \emph{International Conference on
  Machine Learning}, 2013, pp. 1373--1381.

\bibitem{fitzgerald2012vocal}
D.~FitzGerald, ``Vocal separation using nearest neighbours and median
  filtering,'' 2012.

\bibitem{kim2015efficient}
M.~Kim, P.~Smaragdis, and G.~J. Mysore, ``Efficient manifold preserving audio
  source separation using locality sensitive hashing,'' in \emph{2015 IEEE
  International Conference on Acoustics, Speech and Signal Processing
  (ICASSP)}.\hskip 1em plus 0.5em minus 0.4em\relax IEEE, 2015, pp. 479--483.

\bibitem{guo2018bitwise}
L.~Guo and M.~Kim, ``Bitwise source separation on hashed spectra: An efficient
  posterior estimation scheme using partial rank order metrics,'' in \emph{2018
  IEEE International Conference on Acoustics, Speech and Signal Processing
  (ICASSP)}.\hskip 1em plus 0.5em minus 0.4em\relax IEEE, 2018, pp. 761--765.

\bibitem{rickard2002approximate}
S.~Rickard and O.~Yilmaz, ``On the approximate w-disjoint orthogonality of
  speech,'' in \emph{2002 IEEE International Conference on Acoustics, Speech,
  and Signal Processing}, vol.~1.\hskip 1em plus 0.5em minus 0.4em\relax IEEE,
  2002, pp. I--529.

\bibitem{gionis1999similarity}
A.~Gionis, P.~Indyk, R.~Motwani, \emph{et~al.}, ``Similarity search in high
  dimensions via hashing,'' in \emph{Vldb}, vol.~99, no.~6, 1999, pp. 518--529.

\bibitem{mikolov2013distributed}
T.~Mikolov, I.~Sutskever, K.~Chen, G.~S. Corrado, and J.~Dean, ``Distributed
  representations of words and phrases and their compositionality,'' in
  \emph{Advances in neural information processing systems}, 2013, pp.
  3111--3119.

\bibitem{pennington2014glove}
J.~Pennington, R.~Socher, and C.~Manning, ``Glove: Global vectors for word
  representation,'' in \emph{Proceedings of the 2014 conference on empirical
  methods in natural language processing (EMNLP)}, 2014, pp. 1532--1543.

\bibitem{bromley1994signature}
J.~Bromley, I.~Guyon, Y.~LeCun, E.~S{\"a}ckinger, and R.~Shah, ``Signature
  verification using a" siamese" time delay neural network,'' in \emph{Advances
  in neural information processing systems}, 1994, pp. 737--744.

\bibitem{bach2006learning}
F.~R. Bach and M.~I. Jordan, ``Learning spectral clustering, with application
  to speech separation,'' \emph{Journal of Machine Learning Research}, vol.~7,
  no. Oct, pp. 1963--2001, 2006.

\bibitem{cooke2001auditory}
M.~Cooke and D.~P. Ellis, ``The auditory organization of speech and other
  sources in listeners and computational models,'' \emph{Speech communication},
  vol.~35, no. 3-4, pp. 141--177, 2001.

\bibitem{hershey2016deep}
J.~R. Hershey, Z.~Chen, J.~Le~Roux, and S.~Watanabe, ``Deep clustering:
  Discriminative embeddings for segmentation and separation,'' in \emph{2016
  IEEE International Conference on Acoustics, Speech and Signal Processing
  (ICASSP)}.\hskip 1em plus 0.5em minus 0.4em\relax IEEE, 2016, pp. 31--35.

\bibitem{duan2012online}
Z.~Duan, G.~J. Mysore, and P.~Smaragdis, ``Online {PLCA} for real-time
  semi-supervised source separation,'' in \emph{International Conference on
  Latent Variable Analysis and Signal Separation}.\hskip 1em plus 0.5em minus
  0.4em\relax Springer, 2012, pp. 34--41.

\bibitem{vincent2006performance}
E.~Vincent, R.~Gribonval, and C.~F{\'e}votte, ``Performance measurement in
  blind audio source separation,'' \emph{IEEE transactions on audio, speech,
  and language processing}, vol.~14, no.~4, pp. 1462--1469, 2006.

\bibitem{taal2010short}
C.~H. Taal, R.~C. Hendriks, R.~Heusdens, and J.~Jensen, ``A short-time
  objective intelligibility measure for time-frequency weighted noisy speech,''
  in \emph{2010 IEEE International Conference on Acoustics, Speech and Signal
  Processing}.\hskip 1em plus 0.5em minus 0.4em\relax IEEE, 2010, pp.
  4214--4217.

\bibitem{sun2013universal}
D.~L. Sun and G.~J. Mysore, ``Universal speech models for speaker independent
  single channel source separation,'' in \emph{2013 IEEE International
  Conference on Acoustics, Speech and Signal Processing}.\hskip 1em plus 0.5em
  minus 0.4em\relax IEEE, 2013, pp. 141--145.

\end{thebibliography}
%
% or list them by yourself
% \begin{thebibliography}{9}
% 
% \bibitem{waspaa19web}
%   \url{http://www.waspaa.com}.
%
% \bibitem{IEEEPDFSpec}
%   {PDF} specification for {IEEE} {X}plore$^{\textregistered}$,
%   \url{http://www.ieee.org/portal/cms_docs/pubs/confstandards/pdfs/IEEE-PDF-SpecV401.pdf}.
%
% \bibitem{PDFOpenSourceTools}
%   Creating high resolution {PDF} files for book production with 
%   open source tools, 
%   \url{http://www.grassbook.org/neteler/highres_pdf.html}.
%
% \bibitem{eWilliams1999}
% E. Williams, \emph{Fourier Acoustics: Sound Radiation and Nearfield Acoustic
%   Holography}. London, UK: Academic Press, 1999.
% 
% \bibitem{ieeecopyright}
%   \url{http://www.ieee.org/web/publications/rights/copyrightmain.html}.
%
% \bibitem{cJones2003}
% C. Jones, A. Smith, and E. Roberts, ``A sample paper in conference
%   proceedings,'' in \emph{Proc. IEEE ICASSP}, vol. II, 2003, pp. 803--806.
% 
% \bibitem{aSmith2000}
% A. Smith, C. Jones, and E. Roberts, ``A sample paper in journals,'' 
%   \emph{IEEE Trans. Signal Process.}, vol. 62, pp. 291--294, Jan. 2000.
% 
% \end{thebibliography}

\end{sloppy}
\end{document}